\documentclass[letterpaper,12pt]{article}

\usepackage{fullpage}
\usepackage{graphicx}
\usepackage{amsmath}  
\usepackage{amsfonts} 

\begin{document}

\title{Quasi light fields:\\ extending the light field to coherent radiation}

\author{Anthony Accardi and Gregory Wornell \\ {\normalsize Massachusetts Institute of Technology}}
\date{July 30, 2009}

\maketitle

\hskip.3in
\begin{minipage}{5.5in} 
  \parindent.2in\noindent \rm
 Imaging technologies such as dynamic viewpoint generation are
  engineered for incoherent radiation using the traditional light
  field, and for coherent radiation using electromagnetic field
  theory.  We present a model of coherent image formation that
  strikes a balance between the utility of the light field and the
  comprehensive predictive power of Maxwell's equations.  We
  synthesize research in optics and signal processing to formulate,
  capture, and form images from {\em quasi light fields}, which extend
  the light field from incoherent to coherent radiation.  Our coherent
  cameras generalize the classic beamforming algorithm in sensor array
  processing, and invite further research on alternative notions of
  image formation.
  \hfil 
\end{minipage}
\vskip.25in

\section{Introduction}

The light field represents radiance as a function of position and
direction, thereby decomposing optical power flow along rays. The
light field is an important tool used in many imaging applications in
different disciplines, but is traditionally limited to incoherent
light. In computer graphics, a rendering pipeline can compute new
views at arbitrary camera positions from the light field
\cite{LEVO96}. In computational photography, a camera can measure the
light field and later generate images focused at different depths,
after the picture is taken \cite{NG05}. In electronic displays, an
array of projectors can present multiple viewpoints encoded in the
light field, enabling 3D television \cite{CHUN09}. Many recent
incoherent imaging innovations have been made possible by expressing
image pixel values as appropriate integrals over light field rays.

For coherent imaging applications, the value of decomposing power by
position and direction has long been recognized without the aid of a
light field, since the complex-valued scalar field encodes direction
in its phase. A hologram encodes multiple viewpoints, but in a
different way than the light field \cite{ZIEG07}. An ultrasound
machine generates images focused at different depths, but from air
pressure instead of light field measurements \cite{SZAB04}. A Wigner
distribution function models the operation of optical systems in
simple ways, by conveniently inferring direction from the scalar
field instead of computing non-negative light field values
\cite{BAST97}.   Comparing these applications, coherent imaging uses
the scalar field to achieve results similar to those that incoherent
imaging obtains with the light field.

Our goal is to provide a model of coherent image formation that
combines the utility of the light field with the comprehensive
predictive power of the
scalar field. The similarities between coherent and incoherent imaging
motivate exploring how the scalar field and light field are related,
which we address by synthesizing research across three different
communities. Each community is concerned with a particular Fourier
transform pair and has its own name for the light field. In optics,
the pair is position and direction, and Walther discovered the first
generalized radiance function by matching power predictions made with
radiometry and scalar field theory \cite{WALT68}. In quantum physics,
the pair is position and momentum, and Wigner discovered the first
quasi-probability distribution, or phase-space distribution, as an aid
to computing the expectation value of a quantum operator
\cite{WIGN32}. In signal processing, the pair is time and frequency,
and while instantaneous spectra were used as early as 1890 by
Sommerfeld, Ville is generally credited with discovering the first nontrivial
quadratic time-frequency distribution by considering how to distribute
the energy of a signal over time and frequency \cite{VILL48}. Walther,
Wigner, and Ville independently arrived at essentially the same
function, which is one of the ways to express a light field for
coherent radiation in terms of the scalar field.

The light field has its roots in radiometry, a phenomenological theory
of radiative power transport that began with Herschel's observations
of the sun \cite{STEP05}, developed through the work of
astrophysicists such as Chandrasekhar \cite{CHAN60}, and culminated
with its grounding in electromagnetic field theory by Friberg et al.
\cite{FRIB92}. The light field represents radiance, which is the
fundamental quantity in radiometry, defined as power per unit
projected area per unit solid angle. Illuminating engineers would
integrate radiance to compute power quantities, although no one could
validate these calculations with the electromagnetic field theory
formulated by Maxwell. Gershun was one of many physicists who
attempted to physically justify radiometry, and who introduced the
phrase {\em light field} to represent a three-dimensional vector field
analogous to the electric and magnetic fields \cite{GERS36}. Gershun's
light field is a degenerate version of the one we discuss, and more
closely resembles the time-averaged Poynting vector that appears in a
rigorous derivation of geometric optics \cite{BORN99}. Subsequently,
Walther generalized radiometry to coherent radiation in two different
ways\cite{WALT68,WALT73}, and Wolf connected Walther's work to quantum
physics \cite{WOLF78}, ultimately leading to the discovery of many
more generalized radiance functions \cite{AGAR87} and a firm
foundation for radiometry \cite{FRIB92}.

Meanwhile, machine vision researchers desired a representation for all
the possible pictures a pinhole camera might take in space-time, which
led to the current formulation of the light field.  Inspired by
Leonardo da Vinci, Adelson and Bergen defined a plenoptic function to
describe ``everything that can be seen'' as the intensity recorded by
a pinhole camera, parametrized by position, direction, time, and
wavelength \cite{ADEL91}.  Levoy and Hanrahan tied the plenoptic
function more firmly to radiometry, by redefining Gershun's phrase
{\em light field} to mean radiance parametrized by position and
direction \cite{LEVO96}.  Gortler et al. introduced the same
construct, but instead called it the lumigraph \cite{GORT96}.  The
{\em light field} is now the dominant terminology used in incoherent
imaging contexts.

Our contribution is to describe and characterize all the ways to
extend the light field to coherent radiation, and to interpret
coherent image formation using the resulting extended light fields. We
call our extended light fields {\em quasi light fields}, which are analogous
to the generalized radiance functions of optics, the quasi-probability
and phase-space distributions of quantum physics, and the quadratic
class of time-frequency distributions of signal processing. Agarwal et
al. have already extended the light field to coherent radiation
\cite{AGAR87}, and the signal processing community has already
classified all of the ways to distribute power over time and frequency
\cite{BOAS03}. Both have traced their roots to quantum physics. But to
our knowledge, no one has connected the research to show 1) that the
quasi light fields represent {\em all} the ways to extend the light
field to coherent radiation, and 2) that the signal processing
classification informs which quasi light field to use for a specific
application.  We further contextualize the references, making any
unfamiliar literature more accessible to specialists in other areas.

Our paper is organized as follows.  We describe the traditional light
field in Section \ref{sec-tradLF}.  We formulate the quasi light
fields in Section \ref{sec-formulate}, by reviewing and relating the
relevant research in optics, quantum physics, and signal processing.
In Section \ref{sec-capture}, we describe how to capture quasi light
fields, discuss practical sampling issues, and illustrate the impact
of light field choice on energy localization.  In Section \ref{sec-formation},
we describe how to form images with quasi light fields.  We derive a
light field camera, demonstrate and compensate for diffraction
limitations in the near zone, and generalize the classic beamforming
algorithm in sensor array processing.  We conclude the paper in Section
\ref{sec-discussion}, where we remark on the utility of quasi light
fields and future perspectives on image formation.

\section{The Traditional Light Field}
\label{sec-tradLF}

\begin{figure}[ht]
 \centerline{\includegraphics[width=10.4cm]{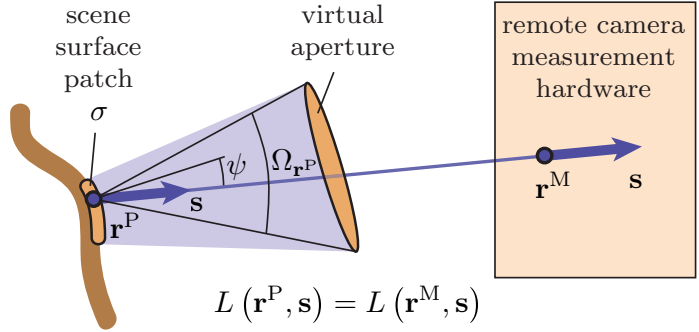}}
 \caption{\small We can compute the value of each pixel in an image produced by an
  arbitrary virtual camera, defined as the power emitted from a
  scene surface patch towards a virtual aperture,
  by integrating an appropriate bundle of light field rays that have
  been previously captured with remote hardware.}
\label{fig-tradradio}
\end{figure}

The light field is a useful tool for incoherent imaging because it
acts as an intermediary between the camera and the picture, decoupling
information capture and image production: the camera measures the
light field, from which many different traditional pictures can be
computed.  We define a pixel in the image of a scene by a surface
patch $\sigma$ and a virtual aperture (Figure \ref{fig-tradradio}).
Specifically, we define the pixel value as the power $P$ radiated by
$\sigma$ towards the aperture, just as an ideal single-lens camera
would measure.  According to radiometry, $P$ is an integral over a
bundle of light field rays \cite{BOYD83}:
\begin{equation}
  P = \int_{\sigma}\int_{\Omega_\mathbf{r}} L(\mathbf{r},\mathbf{s}) \cos\psi\,\textrm{d}^2s\,\textrm{d}^2r,
  \label{eq-radlaw}
\end{equation}
where $L(\mathbf{r},\mathbf{s})$ is the radiance at position
$\mathbf{r}$ and in unit direction $\mathbf{s}$, $\psi$ is the angle
that $\mathbf{s}$ makes with the surface normal at $\mathbf{r}$, and
$\Omega_\mathbf{r}$ is the solid angle subtended by the virtual
aperture at $\mathbf{r}$.  The images produced by many different
conventional cameras can be computed from the light field using
(\ref{eq-radlaw}) \cite{ADAM07}.  

The light field has an important property that allows us to measure it
remotely: the light field is constant along rays in a lossless medium
\cite{BOYD83}.  To measure the light field on the surface of a scene,
we follow the rays for the images we are interested in, and
intercept those rays with our camera hardware (Figure
\ref{fig-tradradio}).  However, our hardware must be capable of
measuring the radiance at a point and in a specific direction; a
conventional camera that simply measures the irradiance at a point is
insufficient.  We can discern directional power flow
using a lens array, as is done in a plenoptic camera \cite{NG05}.

In order to generate coherent images using the same framework
described above, we must overcome three challenges.  First, we must
determine how to measure power flow by position and direction to
formulate a coherent light field.  Second, we must capture the
coherent light field remotely and be able to infer behavior at the
scene surface.  Third, we must be able to use (\ref{eq-radlaw}) to
produce correct power values, so that we can form images by
integrating over the coherent light field.  We address each challenge
in a subsequent section.

\section{Formulating Quasi Light Fields}
\label{sec-formulate}

We motivate, systematically generate, and characterize the
quasi light fields by relating existing research.  We begin in Section
\ref{sec-optics} with research in optics that frames the challenge of
extending the light field to coherent radiation in terms of satisfying
a power constraint required for radiometry to make power predictions
consistent with scalar field theory.  While useful in developing an
intuition for quasi light fields, the power constraint does
not allow us to easily determine the quasi light fields.  We therefore
proceed in Section \ref{sec-quantum} to describe research in quantum
physics that systematically generates quasi light fields satisfying
the power constraint, and that shows how the quasi light fields are
true extensions that reduce to the traditional light field under
certain conditions.  While useful for generating the quasi light
fields, the quantum physics approach does not allow us to easily
characterize them.  Therefore, in Section \ref{sec-signal} we map the
generated quasi light fields to the quadratic class of time-frequency
distributions, which has been extensively characterized and classified
by the signal processing community.  By relating research in optics,
quantum physics, and signal processing, we express all the ways to
extend the light field to coherent radiation, and provide insight on
how to select an appropriate quasi light field for a particular
application.

We assume a perfectly coherent complex scalar field $U(\mathbf{r})$ at
a fixed frequency $\nu$ for simplicity, although we comment in Section
\ref{sec-discussion} on how to extend the results to broadband, partially
coherent radiation. The radiometric theory we discuss assumes a planar
source at $z = 0$. Consequently, although the light field is defined
in three-dimensional space, much of our analysis is confined to planes
$z = z_0$ parallel to the source. Therefore, for convenience, we use
$\mathbf{r} = (x,y,z)$ and $\mathbf{s} = (s_x,s_y,s_z)$ to indicate
three-dimensional vectors and $\mathbf{r}_\perp = (x,y)$ and
$\mathbf{s}_\perp = (s_x,s_y)$ to indicate two-dimensional, projected
versions.

\subsection{Intuition from Optics}
\label{sec-optics}

An extended light field must produce accurate power transport
predictions consistent with rigorous theory; thus the power computed
from the scalar field using wave optics determines the allowable light
fields via the laws of radiometry.  One way to find extended light
fields is to guess a light field equation that satisfies this power
constraint, which is how Walther identified the first extended light
field \cite{WALT68}.  The scenario involves a planar source at $z = 0$
described by $U(\mathbf{r})$, and a sphere of large radius $\rho$
centered at the origin.  We use scalar field theory to compute the
flux through part of the sphere, and then use the definition of
radiance to determine the light field from the flux.

According to scalar field theory, the differential flux $\textrm{d}\Phi$
through a portion of the sphere subtending differential solid angle
$\textrm{d}\Omega$ is given by integrating the radial component of the
energy flux density vector $\mathbf{F}$.  From diffraction theory, the
scalar field in the far zone is 
\begin{equation}
  U^\infty(\rho\mathbf{s}) = -\frac{2\pi i}{k}s_z\frac{\exp(ik\rho)}{\rho}a(\mathbf{s})
  \label{eq-farzone}
\end{equation}
where $k = 2\pi/\lambda$ is the wave number, $\lambda$
is the wavelength, and 
\begin{equation}
  a(\mathbf{s}) = \left(\frac{k}{2\pi}\right)^2 \int U(\mathbf{r})\exp(-ik\mathbf{s}\cdot\mathbf{r})\,\textrm{d}^2r
  \label{eq-planewave}
\end{equation}
is the plane wave component in direction $\mathbf{s}$ \cite{GOOD68}.  Now
\begin{equation}
  \mathbf{F}^\infty(\rho\mathbf{s}) = \left(\frac{2\pi}{k}\right)^2 a(\mathbf{s})a^\ast(\mathbf{s})\frac{s_z^2}{\rho^2}\mathbf{s},
\end{equation}
so that 
\begin{equation}
  \textrm{d}\Phi = \left(\frac{2\pi}{k}\right)^2 s_z^2 a(\mathbf{s})a^\ast(\mathbf{s})\,\textrm{d}\Omega.
\end{equation}

According to radiometry, radiant intensity is flux per unit solid angle
\begin{equation}
  I(\mathbf{s}) = \frac{\textrm{d}\Phi}{\textrm{d}\Omega} = \left(\frac{2\pi}{k}\right)^2 s_z^2 a(\mathbf{s})a^\ast(\mathbf{s}).
  \label{eq-radint}
\end{equation}
Radiance is $I(\mathbf{s})$ per unit projected area \cite{BOYD83}, and this
is where the guessing happens: there are many ways to distribute
(\ref{eq-radint}) over projected area by factoring out $s_z$ and an outer
integral over the source plane, but none yield light fields that satisfy
all the traditional properties of radiance \cite{FRIB79}.  One way to
factor (\ref{eq-radint}) is to substitute the expression for $a(\mathbf{s})$
from (\ref{eq-planewave}) into (\ref{eq-radint}) and change variables:
\begin{equation}
  I(\mathbf{s}) = s_z \int \left[\left(\frac{k}{2\pi}\right)^2s_z \int
    U\left(\mathbf{r}+\frac{1}{2}\mathbf{r}^\prime\right) 
    U^\ast\left(\mathbf{r}-\frac{1}{2}\mathbf{r}^\prime\right)
    \exp\left(-ik\mathbf{s}_\perp\cdot\mathbf{r}_\perp^\prime\right)
    \,\textrm{d}^2r^\prime \right]\,\textrm{d}^2r.
  \label{eq-factored}
\end{equation}
The bracketed expression is Walther's first extended light field
\begin{equation}
  L^\textrm{W}(\mathbf{r},\mathbf{s}) = \left(\frac{k}{2\pi}\right)^2 s_z \mathfrak{W}(\mathbf{r},\mathbf{s}_\perp/\lambda),
  \label{eq-wigrad}
\end{equation}
where
\begin{equation}
  \mathfrak{W}(\mathbf{r},\mathbf{s}_\perp) = \int
  U\left(\mathbf{r}+\frac{1}{2}\mathbf{r}^\prime\right)
  U^\ast\left(\mathbf{r}-\frac{1}{2}\mathbf{r}^\prime\right)
  \exp\left(-i2\pi \mathbf{s}_\perp\cdot \mathbf{r}_\perp^\prime\right)
  \,\textrm{d}^2r^\prime
  \label{eq-wigner}
\end{equation}
is the Wigner distribution \cite{FLAN99}.  We may manually factor
(\ref{eq-radint}) differently to obtain other extended light fields in
an {\em ad hoc} manner, but it is hard to find and verify the
properties of all extended light fields this way, and we would have to
individually analyze each light field that we do manage to find.  So
instead, we pursue a systematic approach to exhaustively identify and
characterize the extended light fields that guarantee the correct
radiant intensity in (\ref{eq-radint}).

\subsection{Generating Explicit Extensions from Quantum Physics}
\label{sec-quantum}

The mathematics of quantum physics provides us with a systematic
extended light field generator that factors the radiant intensity in
(\ref{eq-radint}) in a structured way. Walther's extended light field
in (\ref{eq-wigrad}) provides the hint for this connection
between radiometry and quantum physics. Specifically, Wolf recognized
the similarity between Walther's light field and the Wigner
phase-space distribution \cite{WIGN32} from quantum physics
\cite{WOLF78}. Subsequently, Agarwal, Foley, and Wolf repurposed the
mathematics behind phase-space representation theory to generate new
light fields instead of distributions \cite{AGAR87}. We summarize
their approach, define the class of quasi light fields, describe how
quasi light fields extend traditional radiometry, and show how
quasi light fields can be conveniently expressed as filtered Wigner
distributions.

Agarwal et al.'s key insight was to introduce a position operator
$\hat{\mathbf{r}}_\perp$ and a direction operator $\hat{\mathbf{s}}_\perp$ that obey
the commutation relations \cite{GRIF05}
\begin{equation} 
[\hat{x},\hat{s}_x] = i\lambda / 2\pi, \quad [\hat{y},\hat{s}_y] = i\lambda / 2\pi, 
\label{eq-commute} 
\end{equation} 
and to map the different ways of ordering the operators to different
extended light fields. This formulation is valuable for two reasons.
First, (\ref{eq-commute}) is analogous to the quantum-mechanical
relations for position and momentum, allowing us to exploit the
phase-space distribution generator from quantum physics for our own
purposes, thereby providing an explicit formula for extended light
fields. Second, in the geometric optics limit as $\lambda \to 0$, the
operators commute per (\ref{eq-commute}), so that all of the extended
light fields collapse to the same function that can be related to the
traditional light field. Therefore, Agarwal et al.'s formulation not
only provides us with different ways of expressing the light field for
coherent radiation, but also explains how these differences arise as
the wavelength becomes non-negligible.

We now summarize the phase-space representation calculus that Agarwal
and Wolf invented \cite{AGAR70a} to map operator orderings to
functions, which Agarwal et al.  later applied to radiometry
\cite{AGAR87}, culminating in a formula for extended light fields.
The phase-space representation theory generates a function
$\tilde{L}^\Omega$ from any operator $\hat{L}$ for each distinct way
$\Omega$ of ordering collections of $\hat{\mathbf{r}}_\perp$ and
$\hat{\mathbf{s}}_\perp$.  So by choosing a specific $\hat{L}$ defined
by its matrix elements using the Dirac notation \cite{GRIF05}
\begin{equation} 
 \left\langle \mathbf{r}_\perp^\textrm{R} \big\vert 
\hat{L} 
\big\vert \mathbf{r}_\perp^\textrm{C} \right\rangle 
= U\left(\mathbf{r}^\textrm{R}\right)U^\ast\left(\mathbf{r}^\textrm{C}\right), 
\end{equation} 
and supplying $\hat{L}$ as input, we obtain the extended light fields 
\begin{equation}
 L^\Omega(\mathbf{r},\mathbf{s}) = \left(\frac{k}{2\pi}\right)^2 s_z \tilde{L}^\Omega\left(\mathbf{r}_\perp,\mathbf{s}_\perp\right) 
\end{equation} 
as outputs.  The power constraint from Section \ref{sec-optics}
translates to a minor constraint on the allowed orderings $\Omega$, so
that $L^\Omega$ can be factored from (\ref{eq-radint}).  Finally,
there is an explicit formula for $L^\Omega$ \cite{AGAR70a}, which
in Friberg et al.'s form \cite{FRIB92} reads
\begin{eqnarray}
  L^\Omega(\mathbf{r},\mathbf{s}) &=& \frac{k^2}{(2\pi)^4} s_z  \iiint
  \tilde{\Omega}\left(\mathbf{u}, k\mathbf{r}_\perp^{\prime\prime}\right)
  \exp\left[-i\mathbf{u} \cdot \left(\mathbf{r}_\perp-\mathbf{r}_\perp^\prime\right) \right]
  \exp\left(-ik\mathbf{s}_\perp\cdot\mathbf{r}_\perp^{\prime\prime} \right)
  \nonumber \\ 
  && \times U\left(\mathbf{r}^\prime+\frac{1}{2}\mathbf{r}^{\prime\prime}\right)
  U^\ast\left(\mathbf{r}^\prime-\frac{1}{2}\mathbf{r}^{\prime\prime}\right)
  \,\textrm{d}^2u\,\textrm{d}^2r^\prime\,\textrm{d}^2r^{\prime\prime}, 
  \label{eq-quasiform}
\end{eqnarray}
where $\tilde{\Omega}$ is a functional representation of the ordering $\Omega$.  

Previous research has related the extended light fields $L^\Omega$ to
the traditional light field, by examining how the $L^\Omega$ behave
for globally incoherent light of a small wavelength, an environment
technically modeled by a quasi-homogeneous source in the
geometric optics limit where $\lambda\to 0$.  As $\lambda\to 0$,
$\hat{\mathbf{r}}_\perp$ and $\hat{\mathbf{s}}_\perp$ commute per
(\ref{eq-commute}), so that all orderings $\Omega$ are equivalent and
all of the extended light fields $L^\Omega$ collapse to the same
function.  Since, in the source plane, Foley and Wolf showed that one
of those light fields behaves like traditional radiance \cite{FOLE85}
for globally incoherent light of a small wavelength, all of the
$L^\Omega$ behave like traditional radiance for globally incoherent
light of a small wavelength.  Furthermore, Friberg et al. showed that
many of the $L^\Omega$ are constant along rays, for globally
incoherent light of a small wavelength \cite{FRIB92}.  The $L^\Omega$
thereby subsume the traditional light field, and {\em globally incoherent
light of a small wavelength} is the environment in which traditional
radiometry holds.

To more easily relate $L^\Omega$ to the signal processing
literature, we conveniently express $L^\Omega$ as a filtered Wigner
distribution.  We introduce a function $\Pi$ and substitute
\begin{equation}
  \tilde{\Omega}(\mathbf{u},\mathbf{v}) =
  \iint
  \Pi(-\mathbf{a},-\mathbf{b})
  \exp\left[-i(\mathbf{a} \cdot \mathbf{u} + \mathbf{b} \cdot \mathbf{v}) \right]
  \,\textrm{d}^2a\,\textrm{d}^2b
\end{equation}
into (\ref{eq-quasiform}), integrate first over $\mathbf{u}$, then over $\mathbf{a}$, and finally
substitute
$\mathbf{b} =\mathbf{s}_\perp^\prime-\mathbf{s}_\perp$: 
\begin{eqnarray}
  L^\Omega(\mathbf{r},\mathbf{s}) 
  &=& \left(\frac{k}{2\pi}\right)^2 s_z  \iint
  \Pi(\mathbf{r}_\perp-\mathbf{r}_\perp^\prime,\mathbf{s}_\perp-\mathbf{s}_\perp^\prime)
  \mathfrak{W}(\mathbf{r}^\prime,\mathbf{s}_\perp^\prime/\lambda)
  \,\textrm{d}^2r^\prime\,\textrm{d}^2s^\prime\nonumber\\
  &=& \left(\frac{k}{2\pi}\right)^2 s_z \;
  \Pi(\mathbf{r}_\perp,\mathbf{s}_\perp) \otimes
  \mathfrak{W}(\mathbf{r},\mathbf{s}_\perp/\lambda).
  \label{eq-convrel}
\end{eqnarray}
The symbol $\otimes$ in (\ref{eq-convrel}) denotes convolution in both
$\mathbf{r}_\perp$ and $\mathbf{s}_\perp$.  Each filter kernel $\Pi$
yields a different light field.  There are only minor restrictions on
$\Pi$, or equivalently on $\tilde{\Omega}$.  Specifically, Agarwal and
Wolf's calculus requires that \cite{AGAR70a}
\begin{equation}
  1/\tilde{\Omega} \textrm{ is an entire analytic function with no zeros on the real component axes}.
  \label{eq-techconst}
\end{equation}
Agarwal et al.'s derivation additionally requires that
\begin{equation}
  \tilde{\Omega}(\mathbf{0},\mathbf{v}) = 1 \textrm{ for all } \mathbf{v}, 
  \label{eq-normconst}
\end{equation}
so that $L^\Omega$ satisfies the laws of radiometry and is consistent
with (\ref{eq-radint}) \cite{AGAR87}.

We call the functions $L^\Omega$, the restricted class of extended
light fields that we have systematically generated, {\em quasi light
  fields}, in recognition of their connection with quasi-probability
distributions in quantum physics.

\subsection{Characterization from Signal Processing}
\label{sec-signal}

Although we have identified the quasi light fields and justified how
they extend the traditional light field, we must still show that we
have found all possible ways to extend the light field to coherent
radiation, and we must indicate how to select a quasi light field for
a specific application. We address both concerns by relating quasi
light fields to bilinear forms of $U$ and $U^\ast$ that are
parameterized by position and direction. First, such bilinear forms
reflect all the different ways to represent the energy distribution of
a complex signal in signal processing, and therefore contain all
possible extended light fields, allowing us to identify any
unaccounted for by quasi light fields. Second, we may use the signal
processing classification of bilinear forms to characterize quasi
light fields and guide the selection of one for an application.

To relate quasi light fields to bilinear forms, we must express the
filtered Wigner distribution in (\ref{eq-convrel}) as a bilinear form.
Towards this end, we first express the filter kernel $\Pi$ in terms of
another function $K$:
\begin{equation}
  \Pi(\mathbf{a},\mathbf{b}) = \int 
  K\left(-\mathbf{a}+\frac{\lambda}{2}\mathbf{v},
    -\mathbf{a}-\frac{\lambda}{2}\mathbf{v}\right)
  \exp(-i2\pi \mathbf{b}\cdot\mathbf{v})
  \,\textrm{d}^2v.
  \label{eq-pidef}
\end{equation}
We substitute (\ref{eq-pidef}) into (\ref{eq-convrel}), integrate
first over $\mathbf{s}_\perp^\prime$, then over $\mathbf{v}$, and
finally substitute
\begin{equation}
  \mathbf{r}^\textrm{R} = \mathbf{r}^\prime + \frac{1}{2}\mathbf{r}^{\prime\prime}, \quad 
  \mathbf{r}^\textrm{C} = \mathbf{r}^\prime - \frac{1}{2}\mathbf{r}^{\prime\prime}
\end{equation}
to express the quasi light field as
\begin{eqnarray}
  L(\mathbf{r},\mathbf{s}) &=& \left(\frac{k}{2\pi}\right)^2 s_z 
  \iint U\left(\mathbf{r}^\textrm{R}\right)
  \Big\{
  K\left(\mathbf{r}_\perp^\textrm{R}-\mathbf{r}_\perp,\mathbf{r}_\perp^\textrm{C}-\mathbf{r}_\perp\right)
  \nonumber \\ 
  && \times \exp\left[-ik\mathbf{s}_\perp\cdot\left(\mathbf{r}_\perp^\textrm{R}-\mathbf{r}_\perp^\textrm{C}\right)\right] 
  \Big\} \;
  U^\ast\left(\mathbf{r}^\textrm{C}\right)  
  \,\textrm{d}^2r^\textrm{R}\,\textrm{d}^2r^\textrm{C}.
  \label{eq-kernel}
\end{eqnarray}
We recognize that (\ref{eq-kernel}) is a bilinear form of $U$ and
$U^\ast$, with kernel indicated by the braces.

The structure of the kernel of the bilinear form in (\ref{eq-kernel})
limits $L$ to a shift-invariant energy distribution.  Specifically,
translating the scalar field in (\ref{eq-kernel}) in position and
direction orthogonal to the $z$-axis according to
\begin{equation}
 U(\mathbf{r}) \to U\left(\mathbf{r}-\mathbf{r}^0\right)
 \exp\left(ik\mathbf{s}_\perp^0 \cdot\mathbf{r}_\perp\right)
\end{equation}
results in a corresponding translation in position and direction in
the light field, after rearranging terms:
\begin{equation}
 L(\mathbf{r},\mathbf{s}) \to L\left(\mathbf{r}-\mathbf{r}^0,\mathbf{s}-\mathbf{s}^0\right).
\end{equation}
Such shift-invariant bilinear forms comprise the {\em quadratic class}
of time-frequency distributions, which is sometimes
misleadingly referred to as Cohen's class \cite{BOAS03}.

The quasi light fields represent {\em all} possible ways of extending
the light field to coherent radiation. This is because any reasonably
defined extended light field must be shift-invariant in position and
direction, as translating and rotating coordinates should modify the
scalar field and light field representations in corresponding ways.
Thus, on the one hand, an extended light field must be a quadratic
time-frequency distribution. On the other hand, (\ref{eq-kernel})
implies that quasi light fields span the entire class of quadratic
time-frequency distributions, apart from the constraints on $\Pi$
described at the end of Section \ref{sec-quantum}.  The constraint in
(\ref{eq-normconst}) is necessary to satisfy the power constraint in
(\ref{eq-radint}), which any extended light field must satisfy. The
remaining constraints in (\ref{eq-techconst}) are technical details
concerning analyticity and the location of zeros; extended light
fields strictly need not satisfy these mild constraints, but the light
fields that are ruled out are well-approximated by light fields that
satisfy them.

We obtain a concrete sensor array processing interpretation of quasi light
fields by grouping the exponentials in (\ref{eq-kernel}) with $U$
instead of $K$:
\begin{eqnarray}
  L(\mathbf{r},\mathbf{s}) &=& \left(\frac{k}{2\pi}\right)^2 s_z 
  \iint 
  \Big\{ U\left(\mathbf{r}^\textrm{R}\right)
   \exp\left[ik\mathbf{s}\cdot\left(\mathbf{r}-\mathbf{r}^\textrm{R}\right)\right]
   \Big\}
   K\left(\mathbf{r}_\perp^\textrm{R}-\mathbf{r}_\perp,\mathbf{r}_\perp^\textrm{C}-\mathbf{r}_\perp\right)
  \nonumber \\ 
  && \times
  \Big\{ U\left(\mathbf{r}^\textrm{C}\right)
    \exp\left[ik\mathbf{s}\cdot\left(\mathbf{r}-\mathbf{r}^\textrm{C}\right)\right]
  \Big\}^\ast
  \,\textrm{d}^2r^\textrm{R}\,\textrm{d}^2r^\textrm{C}.
  \label{eq-filterinterp}
\end{eqnarray}
The integral in (\ref{eq-filterinterp}) is the expected value of the
energy of the output of a spatial filter with impulse response
$\exp(ik\mathbf{s}\cdot\mathbf{r})$ applied to the scalar field, when
using $K$ to estimate the correlation
$E[U(\mathbf{r}^\textrm{R})U^\ast(\mathbf{r}^\textrm{C})]$ by
\begin{equation}
  U\left(\mathbf{r}^\textrm{R}\right)
  K\left(\mathbf{r}_\perp^\textrm{R}-\mathbf{r}_\perp,\mathbf{r}_\perp^\textrm{C}-\mathbf{r}_\perp\right)
  U^\ast\left(\mathbf{r}^\textrm{C}\right).
\end{equation}
That is, the choice of quasi light field corresponds to a choice of how to infer
coherence structure from scalar field measurements. In adaptive
beamforming, the spatial filter $\exp(ik\mathbf{s}\cdot\mathbf{r})$
focuses a sensor array on a particular plane wave component, and $K$
serves a similar role as the covariance matrix taper that gives
rise to design features such as diagonal loading \cite{GUER99}. But
for our purposes, the sensor array processing interpretation in
(\ref{eq-filterinterp}) allows us to cleanly separate the choice of
quasi light field in $K$ from the plane wave focusing in the
exponentials.

Several signal processing books meticulously classify the quadratic
class of time-frequency distributions by their properties, and discuss
distribution design and use for various applications
\cite{FLAN99,BOAS03}. We can use these resources to design quasi light
fields for specific applications. For example, if we desire a light
field with fine directional localization, we may first try the Wigner
quasi light field in (\ref{eq-wigrad}), which is a popular starting
choice. We may then discover that we have too many artifacts from
interfering spatial frequencies, called {\em cross terms}, and therefore
wish to consider a reduced interference quasi light field. We might
try the modified B-distribution, which is a particular reduced
interference quasi light field that has a tunable parameter to
suppress interference. Or, we may decide to design our own quasi light
field in a transformed domain using ambiguity functions. The resulting
tradeoffs can be tailoring to specific application requirements.

\section{Capturing Quasi Light Fields}
\label{sec-capture}

To capture an arbitrary quasi light field, we sample and process the
scalar field. In incoherent imaging, traditional light fields are
typically captured by instead making intensity measurements at a
discrete set of positions and directions, as in done in the plenoptic
camera \cite{NG05}. While it is possible to apply the same technique
to coherent imaging, only a small subset of quasi light fields that
exhibit poor localization properties can be captured this way. In
comparison, all quasi light fields can be computed from the scalar
field, as in (\ref{eq-convrel}). We therefore sample the scalar field
with a discrete set of sensors placed at different positions in space,
and subsequently process the scalar field measurements to compute the
desired quasi light field. We describe the capture process for three
specific quasi light fields in Section \ref{sec-samplescalar}, and
demonstrate the different localization properties of these quasi light
fields via simulation in Section \ref{sec-restrade}.

\subsection{Sampling the Scalar Field}
\label{sec-samplescalar}

\begin{figure}[ht]
 \centerline{\includegraphics[width=10.4cm]{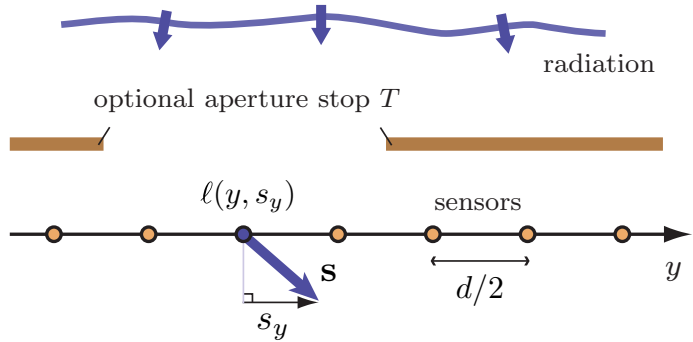}}
 \caption{\small We capture a discrete quasi light field $\ell$ by sampling
 the scalar field at regularly-spaced sensors and processing the
 resulting measurements.  We may optionally apply an aperture stop
 $T$ to mimic traditional light field capture, but this restricts us
 to capturing quasi light fields with poor localization properties.}
\label{fig-samplescalar}
\end{figure}

To make the capture process concrete, we capture three different quasi
light fields. For simplicity, we consider a two-dimensional scene and
sample the scalar field with a linear array of sensors regularly
spaced along the $y$-axis (Figure \ref{fig-samplescalar}). With this
geometry, the scalar field $U$ is parameterized by a single position
variable $y$, and the discrete light field $\ell$ is parameterized by
$y$ and the direction component $s_y$. The sensor spacing is $d/2$,
which we assume is fine enough to ignore aliasing effects. This
assumption is practical for long-wavelength applications such as
millimeter-wave radar. For other applications, aliasing can be avoided
by applying an appropriate pre-filter. From the sensor measurements,
we compute three different quasi light fields, including the
spectrogram and the Wigner.

Although the spectrogram quasi light field is attractive because it
can be captured like a traditional light field by making intensity
measurements, it exhibits poor localization properties. Zhang and
Levoy explain \cite{ZHAN09} how to capture the spectrogram by placing
an aperture stop specified by a transmission function $T$ over the
desired position $y$ before computing a Fourier transform to extract
the plane wave component in the desired direction $s_y$, and
previously Ziegler et al. used the spectrogram as a coherent light
field to represent a hologram \cite{ZIEG07}. The spectrogram is an
important quasi light field because it is the building block for the
quasi light fields that can be directly captured by making intensity
measurements, since all non-negative quadratic time-frequency
distributions, and therefore all non-negative quasi light fields, are sums of
spectrograms \cite{BOAS03}. Ignoring constants and $s_z$, we compute
the discrete spectrogram from the scalar field samples by
\begin{equation}
  \ell^\textrm{S}(y,s_y) = \left| \sum_n T(nd) U(y+nd) \exp\left(-iknds_y\right) \right|^2.
  \label{eq-spectrogramDLF}
\end{equation}

The Wigner quasi light field is a popular choice that exhibits good
energy localization in position and direction \cite{BOAS03}.  We
already identified the Wigner quasi light field in (\ref{eq-wigrad});
the discrete version is
\begin{equation}
  \ell^\textrm{W}(y,s_y) = \sum_n U(y+nd/2)U^\ast(y-nd/2) \exp\left(-iknds_y\right).
  \label{eq-wignerDLF}
\end{equation}
Evidently, the spectrogram and Wigner distribute energy over position
and direction in very different ways.  Per (\ref{eq-spectrogramDLF}),
the spectrogram first uses a Fourier transform to extract directional
information and then computes a quadratic energy quantity, while the
Wigner does the reverse, per (\ref{eq-wignerDLF}).  On the one hand,
this reversal allows the Wigner to better localize energy in position and
direction, since the Wigner is not bound by the classical Fourier uncertainty
principle as the spectrogram is.  On the other hand, the Wigner's
nonlinearities introduce cross-term artifacts by coupling energy in
different directions, thereby replacing the simple uncertainty
principle with a more complicated set of tradeoffs \cite{BOAS03}.  

We now introduce a third quasi light field to capture, in order to
help us understand the implications of requiring quasi light fields to
exhibit traditional light field properties.  Specifically, traditional
light fields have real non-negative values that are zero where the
scalar field is zero, whereas no quasi light field behaves this way
\cite{FRIB79}.  Although the spectrogram has non-negative values, the
support of both the spectrogram and Wigner spills over into regions
where the scalar field is zero.  In contrast, the conjugate Rihaczek
quasi light field, which can be obtained by substituting
(\ref{eq-planewave}) for $a^\ast(\mathbf{s})$ in (\ref{eq-radint}) and
factoring, is identically zero at all positions where the scalar field
is zero and for all directions in which the plane wave component is zero:
\begin{equation}
 L^\textrm{R}(\mathbf{r},\mathbf{s}) = s_z
 U^\ast(\mathbf{r}) 
 \exp(ik\mathbf{s}\cdot\mathbf{r}) a(\mathbf{s}).
 \label{eq-rihrad}
\end{equation}
However, unlike the non-negative spectrogram and the real Wigner, the
Rihaczek is complex-valued, as each of its discoverers independently
observed: Walther in optics \cite{WALT73}, Kirkwood in quantum physics
\cite{KIRK33}, and Rihaczek in signal processing \cite{RIHA68}.  The discrete
conjugate Rihaczek quasi light field is
\begin{equation}
 \ell^\textrm{R}(y,s_y) = U^\ast(y) \exp\left(ikys_y\right) \sum_n U(nd) \exp\left(-iknds_y\right).
 \label{eq-rihaczekDLF}
\end{equation}

\subsection{Localization Tradeoffs}
\label{sec-restrade}

\begin{figure}[ht]
 \centerline{\includegraphics[width=10.4cm]{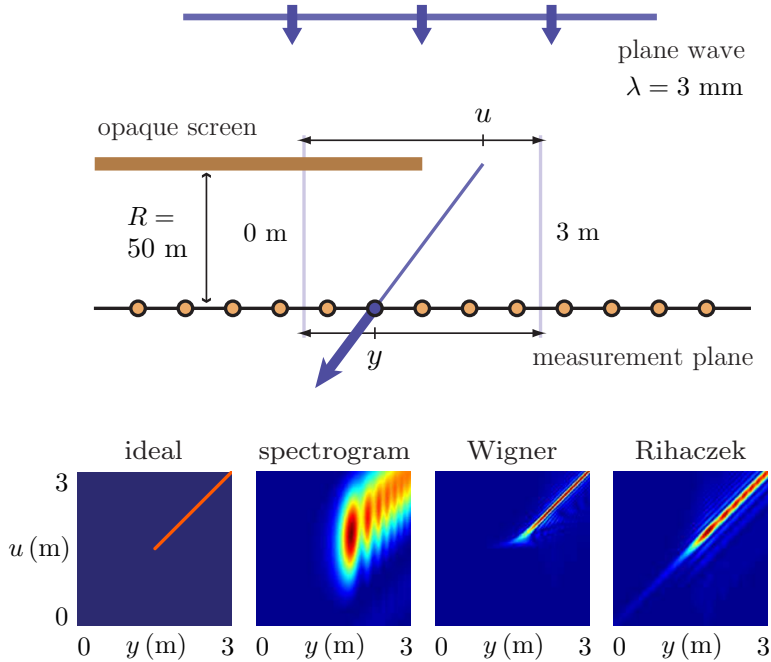}}
 \caption{\small The spectrogram does not resolve a plane wave propagating
   past the edge of an opaque screen as well as other quasi light fields, such
   as the Wigner and Rihaczek.  We capture all three quasi light
   fields by sampling the scalar field with sensors and processing
   the measurements according to (\ref{eq-spectrogramDLF}),
   (\ref{eq-wignerDLF}), and (\ref{eq-rihaczekDLF}).  The ringing
   and blurring in the light field plots indicate the diffraction fringes and
   energy localization limitations.}
 \label{fig-planeedge}
\end{figure}

Different quasi light fields localize energy in position and direction
in different ways, so that the choice of quasi light field impacts the
potential resolution achieved in an imaging application. We illustrate
the diversity of behavior by simulating a plane wave propagating past
a screen edge and computing the spectrogram, Wigner, and Rihaczek
quasi light fields from scalar field samples (Figure
\ref{fig-planeedge}). This simple scenario stresses the main tension
between localization in position and direction: each quasi light field
must encode the position of the screen edge as well as the
downward direction of the plane wave.  The quasi light fields serve as
intermediate representations used to jointly estimate the position of
the screen edge and the orientation of the plane wave.

Our simulation accurately models diffraction using our implementation
of the angular spectrum propagation method, which is the same
technique used in commercial optics software to accurately simulate
wave propagation \cite{ZEMA06}. We propagate a plane wave with
wavelength $\lambda = 3$ mm a distance $R = 50$ m past the screen
edge, where we measure the scalar field and compute the three discrete
light fields using (\ref{eq-spectrogramDLF}), (\ref{eq-wignerDLF}),
and (\ref{eq-rihaczekDLF}). To compute the light fields, we set $d =
\lambda/10$, run the summations over $|n| \leq 10/\lambda$, and use a
rectangular window function of width 10 cm for $T$. We plot
$\ell^\textrm{S}$, $|\ell^\textrm{W}|$, and $|\ell^\textrm{R}|$ in
terms of the two-plane parameterization of the light field
\cite{LEVO96}, so that each ray is directed from a point $u$ in the
plane of the screen towards a point $y$ in the measurement plane, and
so that $s_y = (y - u)/\left[R^2 + (y-u)^2\right]^{1/2}$.
 
We compare each light field's ability to estimate the position of the
screen edge and the orientation of the plane wave (Figure
\ref{fig-planeedge}).  Geometric optics provides an ideal estimate: we
should ideally only see rays pointing straight down ($u = y$) past the
screen edge, corresponding to a diagonal line in the upper-right
quadrant of the light field plots. Instead, we see blurred lines with
ringing. The ringing is physically accurate and indicates the
diffraction fringes formed on the measurement plane. The blurring
indicates localization limitations.  While the spectrogram's window
$T$ can be chosen to narrowly localize energy in either position or
direction, the Wigner narrowly localizes energy in both, depicting
instantaneous frequency without being limited by the classical Fourier
uncertainty principle \cite{BOAS03}.

It may seem that the Wigner light field is preferable to the others
and the clear choice for all applications. While the Wigner light
field possesses excellent localization properties, it exhibits
cross-term artifacts due to interference from different plane wave
components. An alternative quasi light field such as the Rihaczek can
strike a balance between localization and cross-term artifacts, and
therefore may be a more appropriate choice, as discussed at the end of
Section \ref{sec-signal}.  If our goal were to only estimate the
position of the screen edge, we might prefer the spectrogram; to
jointly estimate both position and plane wave orientation, we prefer
the Wigner; and if there were two plane waves instead of one, we might
prefer the Rihaczek.  One thing is certain, however: we must abandon
non-negative quasi light fields to achieve better localization
tradeoffs, as all non-negative quadratic time-frequency distributions
are sums of spectrograms and hence exhibit poor localization tradeoffs
\cite{BOAS03}.

\section{Image Formation}
\label{sec-formation}

We wish to form images from quasi light fields for coherent
applications similarly to how we form images from traditional light
fields for incoherent applications, by using (\ref{eq-radlaw}) to
integrate bundles of light field rays to compute pixel values (Figure
\ref{fig-tradradio}).  However, simply selecting a particular captured
quasi light field $L$ and evaluating (\ref{eq-radlaw}) raises three
questions about the validity of the resulting image.  First,
is it meaningful to distribute coherent energy over surface
area by factoring radiant intensity in (\ref{eq-radint})?  Second,
does the far-zone assumption implicit in radiometry and formalized in
(\ref{eq-farzone}) limit the applicability of quasi field fields?  And
third, how do we capture quasi light field rays remotely if, unlike
the traditional light field, quasi light fields need not be constant
along rays?

The first question is a semantic one. For incoherent light of a small
wavelength, we {\em define} an image in terms of the power radiating
from a scene surface towards an aperture, and physics tells us that
this uniquely specifies the image (Section \ref{sec-formulate}), which
may be expressed in terms of the traditional light field.  If we
attempt to generalize the same definition of an image to partially
coherent, broadband light, and specifically to coherent light at a
non-zero wavelength, we must ask how to isolate the power from a
surface patch towards the aperture, according to classical wave
optics. But there is no unique answer; different isolation techniques
correspond to different quasi light fields. Therefore, to be
well-defined, we must extend the definition of
an image for coherent light to include a particular choice of quasi
light field, which corresponds to a particular factorization of
radiant intensity.

The second and third questions speak of assumptions in the formulation
of quasi light fields and in the image formation from quasi light
fields, which can lead to coherent imaging inaccuracies when these
assumptions are not valid. Specifically, unless the scene surface and
aperture are far apart, the far-zone assumption in (\ref{eq-farzone})
does not hold, so that quasi light fields are incapable of modeling
near-zone behavior. Also, unless we choose a quasi light field that is
constant along rays, such as an angle-impact Wigner function
\cite{ALON01}, remote measurements might not accurately reflect the
light field at the scene surface \cite{LITT93}, resulting in imaging
inaccuracies. Therefore, in general, integrating bundles of remotely
captured quasi light field rays produces an approximation of the image
we have defined. We assess this approximation by building an accurate
near-zone model in Section \ref{sec-physradio}, simulating imaging
performance of several coherent cameras in Section \ref{sec-nearzone},
and showing how our image formation procedure generalizes the classic
beamforming algorithm in Section \ref{sec-genbeam}.

\subsection{Near-Zone Radiometry}
\label{sec-physradio}

\begin{figure}[ht]
 \centerline{\includegraphics[width=10.4cm]{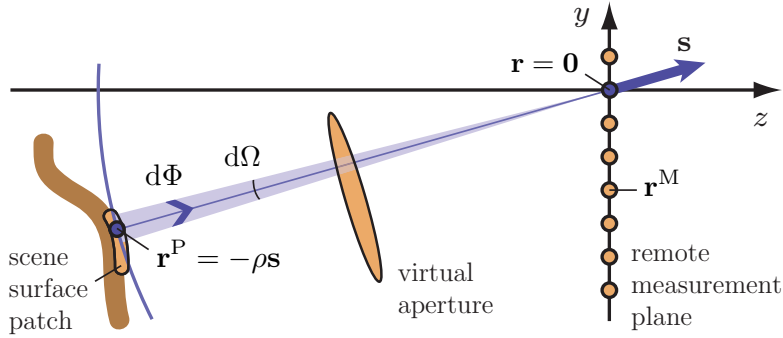}}
 \caption{\small To ensure that integrating bundles of remote light field rays
   in the near zone results in an accurate image, we derive a light field
   $L_\rho^\textrm{R}(\mathbf{r},\mathbf{s})$ in the measurement plane from the
   infinitesimal flux $\textrm{d}\Phi$ at the point $\mathbf{r}^\textrm{P}$
   where the ray originates from the scene surface patch.  We thereby avoid making
   the assumptions that the measurement plane is far from the scene and
   that the light field is constant along rays.}
 \label{fig-physicalradio}
\end{figure}

We take a new approach to formulating light fields for coherent
radiation that avoids making the assumptions that 1) the measurement
plane is far from the scene surface and 2) light fields are constant
along rays. The resulting light fields are accurate in the near zone, and
may be compared with quasi light fields to understand quasi light
field limitations. The key idea is to express a near-zone
light field $L(\mathbf{r},\mathbf{s})$ on the measurement
plane in terms of the infinitesimal flux at the point where the line
containing the ray $(\mathbf{r},\mathbf{s})$ intersects the scene
surface (Figure \ref{fig-physicalradio}). First we compute the scalar
field at the scene surface, next we compute the infinitesimal flux,
and then we identify a light field that predicts the same flux using
the laws of radiometry. In contrast with Walther's approach (Section
\ref{sec-optics}), 1) we do not make the far-zone approximation as in
(\ref{eq-farzone}), and 2) we formulate the light field in the
measurement plane instead of in the source plane at the scene surface.
Therefore, in forming an image from a near-zone light field,
we are not limited to the far zone and we need not relate the light
field at the measurement plane to the light field at the scene
surface.

The first step in deriving a near-zone light field
$L$ for the ray $(\mathbf{r},\mathbf{s})$ is to use the
scalar field on the measurement plane to compute the scalar field at
the point $\mathbf{r}^\textrm{P}$ where the line containing the ray
intersects the scene surface. We choose coordinates so that the
measurement plane is the $xy$-plane, the scene lies many wavelengths
away in the negative $z < 0$ half-space, and $\mathbf{r}$ is at the
origin. We denote the distance between the source
$\mathbf{r}^\textrm{P}$ on the scene surface and the point of
observation $\mathbf{r}$ by $\rho$. Under a reasonable bandwidth
assumption, the inverse diffraction formula expresses the scalar field
at $\mathbf{r}^\textrm{P}$ in terms of the scalar field on the
measurement plane \cite{SHEW68}: \begin{equation}
  U(\mathbf{r}^\textrm{P}) = 
  \frac{ik}{2\pi}\int U(\mathbf{r}^\textrm{M})
  \frac{-z^\textrm{P}}{|\mathbf{r}^\textrm{P}-\mathbf{r}^\textrm{M}|}
  \frac{\exp(-ik|\mathbf{r}^\textrm{P}-\mathbf{r}^\textrm{M}|)}{|\mathbf{r}^\textrm{P}-\mathbf{r}^\textrm{M}|}
  \,\textrm{d}^2r^\textrm{M}.
  \label{eq-inversediff}
\end{equation}

Next, we compute the differential flux $\textrm{d}\Phi$ through a
portion of a sphere at $\mathbf{r}^\textrm{P}$ subtending differential
solid angle $\textrm{d}\Omega$. We obtain $\textrm{d}\Phi$ by
integrating the radial component of the energy flux density vector
\begin{equation}
  \mathbf{F}(\mathbf{r}^\textrm{P}) = -\frac{1}{4\pi k \nu}
  \left[ \frac{\partial U^\ast}{\partial t}\nabla U + \frac{\partial U}{\partial t}\nabla U^\ast \right].
  \label{eq-fluxdensity}
\end{equation}
To keep the calculation simple, we ignore amplitude decay across
the measurement plane,
approximating
\begin{equation}
  |\mathbf{r}^\textrm{P} - \mathbf{r}^\textrm{M}| \approx |\mathbf{r}^\textrm{P}| 
  \label{eq-farapproxdenom}
\end{equation}
outside the exponential in (\ref{eq-inversediff}), and 
\begin{equation}
  \frac{\partial}{\partial |\mathbf{r}^\textrm{P}|} |\mathbf{r}^\textrm{P} -\mathbf{r}^\textrm{M}| \approx 1,
  \label{eq-derapprox}
\end{equation}
when evaluating (\ref{eq-fluxdensity}), resulting in
\begin{equation}
  \mathbf{F}(-\rho\mathbf{s}) =  \left(\frac{2\pi}{k}\right)^2 
  \tilde{a}(-\rho\mathbf{s})  \tilde{a}^\ast(-\rho\mathbf{s}) 
  \frac{s_z^2}{\rho^2}\mathbf{s},
\end{equation}
where
\begin{equation}
  \tilde{a}(-\rho\mathbf{s}) = 
  \left(\frac{k}{2\pi}\right)^2
  \int U(\mathbf{r}^\textrm{M})
  \exp(-ik|-\rho\mathbf{s}-\mathbf{r}^\textrm{M}|)
  \,\textrm{d}^2r^\textrm{M}.
  \label{eq-atilde}
\end{equation}
Thus,
\begin{equation}
  \textrm{d}\Phi = \left(\frac{2\pi}{k}\right)^2 s_z^2 
  \tilde{a}(-\rho\mathbf{s})\tilde{a}^\ast(-\rho\mathbf{s})
  \,\textrm{d}\Omega.
  \label{eq-hardfactor}
\end{equation}

Finally, we factor out $s_z$ and an outer integral over surface area
from $\textrm{d}\Phi/\textrm{d}\Omega$ to determine a near-zone
light field. Unlike in Section \ref{sec-optics}, the
nonlinear exponential argument in $\tilde{a}$ complicates the
factoring. Nonetheless, we obtain a near-zone light field that generalizes the Rihaczek by
substituting (\ref{eq-atilde}) for $\tilde{a}^\ast$ in
(\ref{eq-hardfactor}).  After factoring and freeing $\mathbf{r}$ from
the origin by substituting $\mathbf{r}-\rho\mathbf{s}$ for
$-\rho\mathbf{s}$, we obtain
\begin{eqnarray}
  L_\rho^\textrm{R}(\mathbf{r},\mathbf{s}) &=& 
  s_zU^\ast(\mathbf{r})\exp(ik\rho)\tilde{a}(\mathbf{r}-\rho\mathbf{s})
  \nonumber \\
  &=&
  \left(\frac{k}{2\pi}\right)^2 s_z 
  U^\ast(\mathbf{r}) 
  \exp(ik\rho)
  \int U\left(\mathbf{r}^\textrm{M}\right)
  \exp\left(-ik|\mathbf{r}-\rho\mathbf{s}-\mathbf{r}^\textrm{M}|\right)
  \,\textrm{d}^2r^\textrm{M},
 \label{eq-palight}
\end{eqnarray}
where the subscript $\rho$ reminds us of this near-zone light field's
dependence on distance.

$L_\rho^\textrm{R}$ is evidently neither the traditional
light field nor a quasi light field, as it depends directly on the scene geometry
through an additional distance parameter. This distance
parameter $\rho$ is a function of $\mathbf{r}$, $\mathbf{s}$, and the
geometry of the scene; it is the distance along $\mathbf{s}$ between
the scene surface and $\mathbf{r}$. We may integrate $L_\rho^\textrm{R}$
over a bundle of rays to compute the image pixel values just like any
other light field, as long as we supply the right value of $\rho$ for
each ray.  In contrast, quasi light fields are incapable of modeling optical
propagation in the near zone, as it is insufficient to specify power
flow along rays. We must also know the distance between the source and
point of measurement along each ray.

We can obtain near-zone generalizations of all quasi light fields
through the sensor array processing interpretation in Section
\ref{sec-signal}.  Recall that each quasi light field corresponds to a
particular choice of the function $K$ in (\ref{eq-filterinterp}).  For
example, setting $K(\mathbf{a},\mathbf{b}) = \delta(\mathbf{b})$, where
$\delta$ is the Dirac delta function, yields the Rihaczek quasi light
field $L^\textrm{R}$ in (\ref{eq-rihrad}).  To generalize quasi light
fields to the near zone, we focus at a point instead of a plane wave
component by using a spatial filter with impulse response
$\exp\left(-ik\left|\mathbf{r}-\rho\mathbf{s}\right|\right)$ instead of
$\exp(ik\mathbf{s}\cdot\mathbf{r})$ in (\ref{eq-filterinterp}). Then,
choosing $K(\mathbf{a},\mathbf{b}) = \delta(\mathbf{b})$ yields
$L_\rho^\textrm{R}$, the near-zone generalization of the Rihaczek in
(\ref{eq-palight}), and choosing other functions $K$ yield near-zone
generalizations of the other quasi light fields.

\subsection{Near-Zone Diffraction Limitations}
\label{sec-nearzone}

\begin{figure}[ht] 
 \centerline{\includegraphics[width=10.4cm]{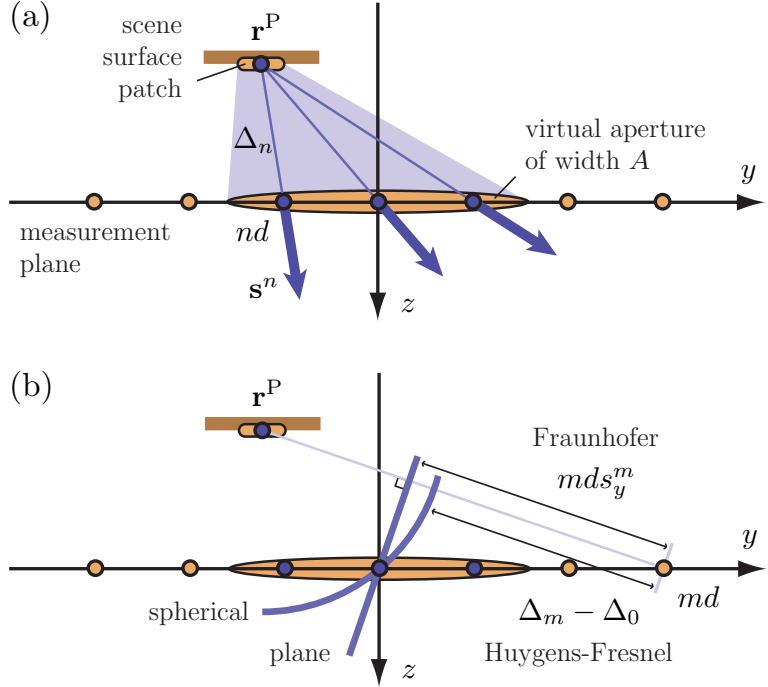}}
 \caption{\small The near-zone
   light field results in a camera that aligns spherical
   wavefronts diverging from the point of focus
   $\mathbf{r}^\textrm{P}$, in accordance with the Huygens-Fresnel
   principle of diffraction, while quasi light fields result in
   cameras that align plane wavefront approximations, in accordance
   with Fraunhofer diffraction. Quasi light fields are therefore only
   accurate in the far zone.  We derive both cameras by approximating
   the integral over a bundle of rays by the summation of discrete
   light field rays (a), and we interpret the operation of each
   camera by how they align sensor measurements along wavefronts from
   $\mathbf{r}^\textrm{P}$ (b).}
\label{fig-planevssphere}
\end{figure}

We compute and compare image pixel values using the Rihaczek quasi
light field $L^\textrm{R}$ and its near-zone generalization
$L_\rho^\textrm{R}$, demonstrating how all quasi light fields implicitly
make the Fraunhofer diffraction approximation that limits accurate
imaging to the far zone. First, we construct coherent cameras from 
$L^\textrm{R}$ and $L_\rho^\textrm{R}$.  For simplicity, we consider a
two-dimensional scene and sample the light fields, approximating the
integral over a bundle of rays (Figure \ref{fig-tradradio}) by the
summation of discrete rays directed from the center
$\mathbf{r}^\textrm{P}$ of the scene surface patch to each sensor on a
virtual aperture of diameter $A$, equally spaced every distance $d$ in
the measurement plane (Figure \ref{fig-planevssphere}a). Ignoring
constants and $s_z$, 
we compute the pixel values for a far-zone camera from the Rihaczek
quasi light field in (\ref{eq-rihrad}), 
\begin{equation}
 P^\textrm{R} =
 \sum_{|nd|<A/2}
   \left[U(nd)  \exp\left(-iknds_y^n\right)\right]^\ast
 \left[\sum_m U(md) \exp\left(-ikmds_y^n\right)\right],
\label{eq-rihacam}
\end{equation}
and for a near-zone camera from the near-zone generalization of the
Rihaczek in (\ref{eq-palight}),
\begin{equation}
P_\rho^\textrm{R} =
\left[\sum_{|nd|<A/2} U(nd) \exp\left(-ik\Delta_n\right)\right]^\ast
\left[\sum_m U(md) \exp\left(-ik\Delta_m\right)\right].
\label{eq-physcam}
\end{equation}
In (\ref{eq-rihacam}), $\mathbf{s}^n$ denotes the unit direction
from $\mathbf{r}^\textrm{P}$ to the $n^\textrm{th}$ sensor,
and in (\ref{eq-physcam}), $\Delta_n$ denotes the distance between
$\mathbf{r}^\textrm{P}$ and the $n^\textrm{th}$ sensor.

By comparing the exponentials in (\ref{eq-rihacam}) with those in
(\ref{eq-physcam}), we see that the near-zone camera aligns the sensor
measurements along spherical wavefronts diverging from the point of
focus $\mathbf{r}^\textrm{P}$, while the far-zone camera aligns
measurements along plane wavefront approximations (Figure
\ref{fig-planevssphere}b). Spherical wavefront alignment makes
physical sense in accordance with the Huygens-Fresnel principle of
diffraction, while approximating spherical wavefronts with plane
wavefronts is reminiscent of Fraunhofer diffraction. In fact, the
far-zone approximation in (\ref{eq-farzone}) used to derive quasi
light fields follows directly from the Rayleigh-Sommerfeld diffraction
integral by linearizing the exponentials, which is precisely
Fraunhofer diffraction. Therefore, all quasi light fields are only
valid for small Fresnel numbers, when the source and point of
measurement are sufficiently far away from each other.

\begin{figure}[ht]
 \centerline{\includegraphics[width=10.4cm]{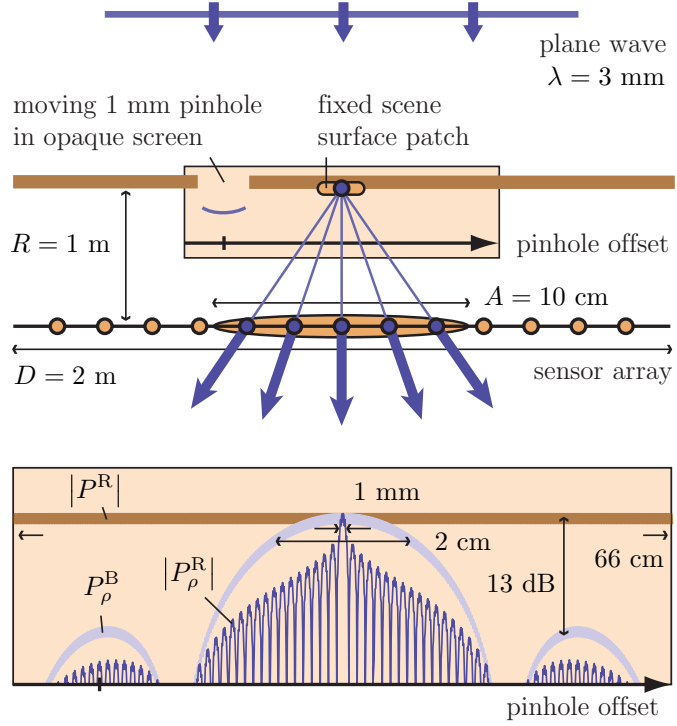}}
 \caption{\small Images of nearby objects formed from pure quasi light fields
   are blurry.  In the scene, a small backlit pinhole moves across the
   field of view of a sensor array that implements three cameras, each computing one pixel
   value for each pinhole position, corresponding to a fixed surface
   patch.  As the pinhole crosses the fixed scene surface patch, the near-zone camera resolves the pinhole down to
   its actual size of 1 mm, while the far-zone camera records a blur 66 cm
   wide.}
 \label{fig-nearzonesim}
\end{figure}

We expect the near-zone camera to outperform the far-zone
camera in near-zone imaging applications, which we demonstrate by
comparing their ability to resolve small targets moving past their
field of view. As a baseline, we introduce a third camera with
non-negative pixel values $P_\rho^\textrm{B}$ by restricting the summation
over $m$ in (\ref{eq-physcam}) to $|md|<A/2$, which results in the
beamformer camera used in sensor array processing \cite{SZAB04,VANT02}.
Alternatively, we could extend the summation over $n$ in
(\ref{eq-physcam}) to the entire array, but this would average
anisotropic responses over a wider aperture diameter, resulting in a
different image.  We simulate an opaque screen containing a pinhole
that is backlit with a coherent plane wave (Figure
\ref{fig-nearzonesim}).  The sensor array is $D = 2$ m wide and just
$R = 1$ m away from the screen. The virtual aperture is $A = 10$ cm
wide and the camera is focused on a fixed 1 mm pixel straight ahead on
the screen. The pinhole has width 1 mm, which is smaller than the
wavelength $\lambda = 3$ mm, so the plane wavefronts bend into
slightly spherical shapes via diffraction. We move the pinhole to the
right, recording pixel values $|P^\textrm{R}|$, $|P_\rho^\textrm{R}|$, and
$P_\rho^\textrm{B}$ for each camera at each pinhole position. Due to
the nature of the coherent combination of the sensor measurements that
produces the pixel values, each camera records a multi-lobed
response. The width of the main lobe indicates the near-zone
resolution of the camera.

The near-zone camera is able to resolve the pinhole down to
its actual size of 1 mm, greatly outperforming the far-zone camera
which records a blur 66 cm wide, and even outperforming the beamformer
camera. Neither comparison is surprising. First, with a Fresnel
number of $D^2/R\lambda \approx 1333$, the Fraunhofer approximation
implicitly made by quasi light fields does not hold for this scenario,
so we expect the far-zone camera to exhibit poor resolution. Second,
the near-zone camera uses the entire $D = 2$ m array instead of
just the sensors on the virtual aperture that the beamformer camera is
restricted to, and the extra sensors lead to improved resolution.

\subsection{Generalized Beamforming}
\label{sec-genbeam}

We compare image formation from light fields with traditional
perspectives on coherent image formation, by relating quasi light
fields and our coherent cameras with the classic beamforming algorithm
used in many coherent imaging applications, including ultrasound
\cite{SZAB04} and radar \cite{VANT02}. The beamforming algorithm
estimates a spherical wave diverging from a point of focus
$\mathbf{r}^\textrm{P}$ by delaying and averaging sensor measurements.
When the radiation is narrowband, the delays are approximated by phase
shifts. With the sensor array geometry from Section
\ref{sec-nearzone}, the beamformer output is 
\begin{equation}
  g = \sum_m T(md) U(md) \exp(-ik\Delta_m),
  \label{eq-beamnear}
\end{equation}
where the $T(md)$ are amplitude weights used to adjust the
beamformer's performance.  As $\mathbf{r}^\textrm{P}$ moves into the
far zone,
\begin{equation}
  \Delta_m - \Delta_0 \to mds_y^m \to mds_y^0,
\end{equation}
so that apart from a constant phase offset, (\ref{eq-beamnear}) becomes a short-time
Fourier transform
\begin{equation}
  g^\infty = \sum_m T(md) U(md) \exp(-ikmds_y^0).
  \label{eq-beamfar}
\end{equation}
Evidently, $|g^\infty|^2$ is a spectrogram quasi light field, and we
may select $T$ to be a narrow window about a point $\mathbf{r}$ to
capture $L^\textrm{S}(\mathbf{r},\mathbf{s}^0)$.  We have already
seen how quasi light fields generalize the spectrogram.

Beamformer applications instead typically select $T$ to be a wide
window to match the desired virtual aperture, and assign the
corresponding pixel value to the output power $|g|^2$.  We can
decompose the three cameras in Section \ref{sec-nearzone} into
such beamformers.  
First, we write $P_\rho^\textrm{R}$ in
(\ref{eq-physcam}) in terms of two different beamformers,
\begin{equation}
  P_\rho^\textrm{R} = g_1^\ast g_2,
\end{equation}
where
\begin{equation}
  g_1 = \sum_{|nd|<A/2} U(nd) \exp\left(-ik\Delta_n\right)
  \label{eq-beamform1}
\end{equation}
and
\begin{equation}
  g_2 = \sum_m U(md) \exp\left(-ik\Delta_m\right),
  \label{eq-beamform2}
\end{equation}
so that the windows for $g_1$ and $g_2$ are rectangular with widths
matching the aperture $A$ and sensor array $D$, respectively.  
Next, by construction 
\begin{equation}
  P_\rho^\textrm{B} = |g_1|^2.
\end{equation}
Finally, in the far zone, $\mathbf{s}^n \to \mathbf{s}^0$ in
(\ref{eq-rihacam}) so that
\begin{equation}
  P^\textrm{R} \to \left(g_1^\infty \right)^\ast g_2^\infty,
\end{equation}
where $g_1^\infty$ and $g_2^\infty$ are given by (\ref{eq-beamfar}) with the
windows $T$ used in (\ref{eq-beamform1}) and (\ref{eq-beamform2}).  
In other words, the near-zone camera is the Hermitian
product of two different beamformers, and is equivalent to the far-zone
camera in the far zone.

We interpret the role of each component beamformer from the
derivation of (\ref{eq-physcam}). Beamformer $g_1^\ast$ aggregates
power contributions across the aperture using measurements of the
conjugate field $U^\ast$ on the aperture, while beamformer $g_2$
isolates power from the point of focus using all available
measurements of the field $U$. In this manner, the tasks of
aggregating and isolating power contributions are cleanly divided
between the two beamformers, and each beamformer uses the measurements
from those sensors appropriate to its task. In contrast, the
beamformer camera uses the same set of sensors for both the power
aggregation and isolation tasks, thereby limiting its ability to
optimize over both tasks.

The near-zone camera achieves a new tradeoff between
resolution and anisotropic sensitivity. We noted that the near-zone
camera exhibits better resolution than the beamformer, for
the same virtual aperture (Figure \ref{fig-nearzonesim}). This is not
an entirely fair comparison because the near-zone camera is
using sensor measurements outside the aperture, and indeed, a
beamformer using the entire array would achieve comparable resolution.
However, extending the aperture to the entire array results in a
different image, as anisotropic responses are averaged over a wider
aperture diameter. We interpret the near-zone camera's
behavior by computing the magnitude 
\begin{equation}
 \left|P_\rho^\textrm{R}\right| = \sqrt{|g_1|^2 |g_2|^2}.
 \label{eq-geomean}
\end{equation}
Evidently, the pixel magnitude of the near-zone camera is
the geometric mean of the two traditional beamformer
output powers.  $|P_\rho^\textrm{R}|$ has better resolution than $|g_1|^2$
and better anisotropic sensitivity than $|g_2|^2$.  

Image formation with alternative light fields uses the conjugate
field and field measurements to aggregate and isolate power in
different ways. In general, image pixel values do not neatly factor
into the product of beamformers, as they do with the Rihaczek.

\section{Concluding Remarks}
\label{sec-discussion}

We enable the use of existing incoherent imaging tools for coherent
imaging applications, by extending the light field to coherent
radiation.  We explain how to formulate, capture, and form images from
quasi light fields.  By synthesizing existing research in optics, quantum
physics, and signal processing, we motivate
quasi light fields, show how quasi light fields extend the traditional
light field, and characterize the properties of different quasi light
fields.  We explain why capturing quasi light fields directly with
intensity measurements is inherently limiting, and demonstrate via
simulation how processing scalar field measurements in different ways
leads to a rich set of energy localization tradeoffs.  We show how
coherent image formation using quasi light fields is complicated by an
implicit far-zone (Fraunhofer) assumption and the fact that not all
quasi light fields are constant along rays.  We demonstrate via
simulation that a pure light field representation is incapable of
modeling near-zone diffraction effects, but that quasi light fields
can be augmented with a distance parameter for greater near-zone
imaging accuracy.  We show how image formation using light
fields generalizes the classic beamforming algorithm, allowing for
new tradeoffs between resolution and anisotropic sensitivity.

Although we have assumed perfectly coherent radiation, tools from
partial coherence theory 1) allow us to generalize our results, and 2) provide
an alternative perspective on image formation. First, our results extend
to broadband radiation of any state of partial coherence by replacing
$U(\mathbf{r}^\textrm{R})U^\ast(\mathbf{r}^\textrm{C})$ with the
cross-spectral density
$W(\mathbf{r}^\textrm{R},\mathbf{r}^\textrm{C},\nu)$. $W$ provides a
statistical description of the radiation, indicating how light at two
different positions, $\mathbf{r}^\textrm{R}$ and
$\mathbf{r}^\textrm{C}$, is correlated at each frequency $\nu$
\cite{MAND95}. Second, $W$ itself may be propagated along rays in an
approximate asymptotic sense \cite{ZYSK05,SCHO08}, which forms the
basis of an entirely different framework for using rays for image
formation, using the cross-spectral density instead of the light field
as the core representation.

We present a model of coherent image formation that strikes a balance
between utility and comprehensive predictive power. On the one hand,
quasi light fields offer more options and tradeoffs than their
traditional, incoherent counterpart. In this manner, the connection
between quasi light fields and quasi-probability distributions in
quantum physics reminds us of the potential benefits of forgoing a
single familiar tool in favor of a multitude of useful yet less
familiar ones. On the other hand, compared with Maxwell's equations,
quasi light fields are less versatile. Therefore, quasi light fields
are attractive to researchers who desire more versatility than
traditional energy-based methods, yet a more specialized model of
image formation than Maxwell's equations.

Quasi light fields illustrate the limitations of the simple definition
of image formation ubiquitous in incoherent imaging. An image is the
visualization of some underlying physical reality, and the energy
emitted from a portion of a scene surface towards a virtual aperture
is not a physically precise quantity when the radiation is coherent,
according to classical electromagnetic wave theory. Perhaps a
different image definition may prove more fundamental for coherent
imaging, or perhaps a quantum optics viewpoint is required for
precision. Although we have borrowed the mathematics from quantum
physics, our entire discussion has been classical. Yet if we introduce
quantum optics and the particle nature of light, we may unambiguously
speak of the probability that a photon emitted from a portion of a
scene surface is intercepted by a virtual aperture. 

\section*{Acknowledgements}

This work was supported, in part, by Microsoft Research, MIT Lincoln
Laboratory, and Semiconductor Research Corporation through the FCRP
Center for Circuit \& System Solutions (C2S2).


\end{document}